\documentclass[manuscript,nonacm]{acmart}
\usepackage{xeCJK}
\setCJKsansfont{FandolHei-Regular.otf}[BoldFont=FandolHei-Bold.otf]
\usepackage{tabularx}
\usepackage{tikz}
\makeatletter\@ACM@balancefalse\makeatother
\AtBeginDocument{%
  }

\begin{document}
\usetikzlibrary{positioning,arrows.meta}

\title[But Can You Draw It Live?]{But Can You Draw It Live? The Emergence of AI Duidus for Adjudicating Suspected Generative AI Use in Digital Art Commission Communities}

\author{Elise Chenxinran Shen}
\email{elise@cs.toronto.edu}
\correspondingauthor
\affiliation{%
  \institution{University of Toronto}
  \department{Computer Science}
  \city{Toronto}
  \state{Ontario}
  \country{Canada}
}

\author{Dashiel Carrera}
\affiliation{%
  \institution{University of Toronto}
  \city{Toronto}
  \state{Ontario}
  \country{Canada}
}

\author{Michelle Annett}
\affiliation{%
  \institution{MishMashMakers}
  \city{Newmarket}
  \state{Ontario}
  \country{Canada}
}

\author{Daniel Wigdor}
\affiliation{%
  \institution{University of Toronto}
  \department{Department of Computer Science}
  \city{Toronto}
  \state{Ontario}
  \country{Canada}
}

\renewcommand{\shortauthors}{Shen et al.}

\begin{abstract}

Generative AI poses new challenges for digital art commission communities, where a commission's value often depends on human labor and authorship. We examine a novel community-developed dispute resolution mechanism called an \textit{AI duidu}, or livestreamed drawing bet (对赌), used when an artist is suspected of using generative AI. In an \textit{AI duidu}, an accuser challenges an artist to create a new artwork live, and community members vote on whether the artist created the disputed artwork without AI assistance. Often, both parties stake money with an intermediary, who releases it after the vote. Through interviews with 17 accusers, accused artists, or intermediaries, we examined how the \textit{AI duidu} emerged and how participants perceived its effectiveness and risks. We found that \textit{AI duidus} shift authorship verification from the artwork itself to the artist, using livestreamed creative labor, community adjudication, and wagering to establish provenance and negotiate community norms surrounding AI use.
\end{abstract}



\maketitle

\section{Introduction}
Digital art commissions (DACs) are bespoke artworks created in response to a commissioner's brief. They have become a substantial form of creative labor for online artists~\cite{porquet2025copying, duffy2017not}. DACs are typically arranged through either dedicated commission platforms or general-purpose social media. Commission platforms include marketplaces such as Fiverr \cite{fiverrterms}, VGen \cite{vgen}, Skeb \cite{skeb2025}, Pixiv\cite{pixiv2025}, and Mihuashi in China~\cite{Mihuashi, zhu2021shequan, wu2026keep}. These platforms support millions of users and facilitate large volumes of commission transactions, illustrating the scale of contemporary commission economies ~\cite{fiverr2026results, skeb2025, pixiv2025}. Alternatively, artists may solicit and negotiate commissions directly with commissioners on platforms such as X (formerly Twitter), Bluesky, Weibo, and Xiaohongshu (RedNote). Unlike platform-mediated commissions, these arrangements rely on public posts and private messages rather than dedicated commission infrastructures.

\begin{figure*}[t]
  \centering
  \includegraphics[width=\textwidth]{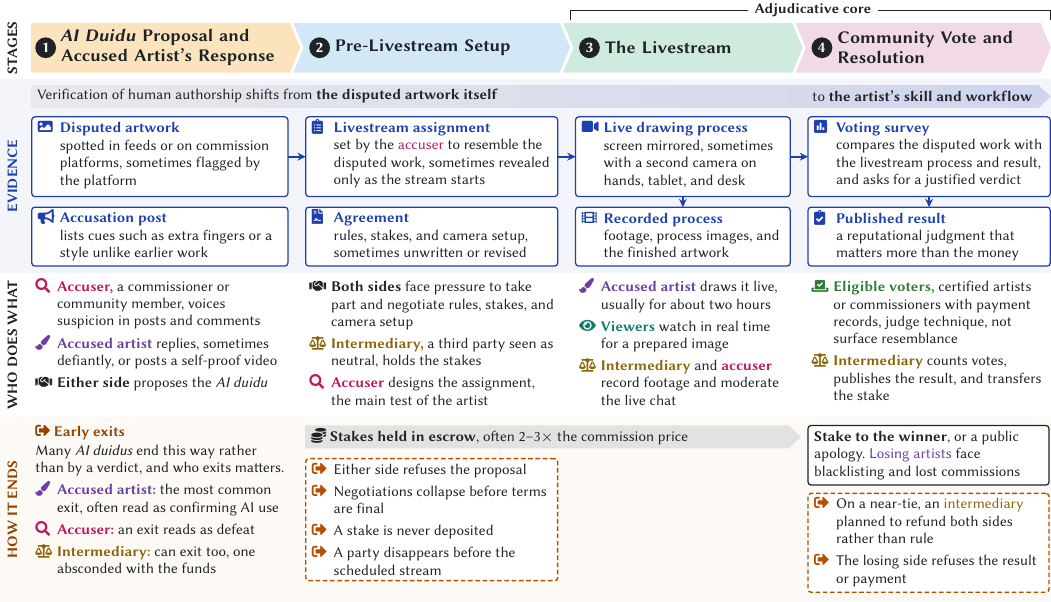}
  \caption{The four stages of an \textit{AI duidu}. For each stage, the rows show the evidence examined, who does what, and how the \textit{AI duidu} ends. Each stage has its own color. A bracket marks Stages 3 and 4 as the adjudicative core, and the band below the stages shows verification shifting from the disputed artwork to the artist's skill and workflow. Dashed boxes list the exits under the stage they leave from. Examples appear in Figures~\ref{fig:p7-post} (Stage 1), \ref{fig:agreement} (Stage 2), \ref{fig:livestream-p11} and \ref{fig:livestream-desk} (Stage 3), and \ref{fig:survey} and \ref{fig:vote-result} (Stage 4).}
  \Description{A four-column diagram of the AI duidu procedure. Across the top, in a row labeled Stages, four chevrons in different colors name the stages: 1, AI Duidu Proposal and Accused Artist's Response, 2, Pre-Livestream Setup, 3, The Livestream, and 4, Community Vote and Resolution. Stages 3 and 4 are bracketed as the adjudicative core. Below them, an arrow reads that verification of human authorship shifts from the disputed artwork itself to the artist's skill and workflow. Three labeled rows follow. The Evidence row has two boxes per stage: the disputed artwork and the accusation post in stage 1, the livestream assignment and the agreement in stage 2, the live drawing process and the recorded process in stage 3, and the voting survey and the published result in stage 4. The Who Does What row marks each role with its own icon and color. In stage 1, the accuser voices suspicion, the accused artist replies, sometimes defiantly or with a self-proof video, and either side proposes the AI duidu. In stage 2, both sides negotiate rules, stakes, and camera setup, an intermediary holds the stakes, and the accuser designs the assignment. In stage 3, the accused artist draws live for about two hours, viewers watch for a prepared image, and the intermediary and accuser record footage and moderate the chat. In stage 4, eligible voters judge technique rather than surface resemblance, and the intermediary counts votes, publishes the result, and transfers the stake. The How It Ends row explains that many AI duidus end through early exits: an accused artist's exit is often read as confirming AI use, an accuser's exit as defeat, and one intermediary absconded with the funds. A band across stages 2 and 3 notes that stakes are held in escrow, often two to three times the commission price. A dashed box under stage 2 lists exits before the livestream: either side refuses the proposal, negotiations collapse, a stake is never deposited, or a party disappears before the stream. Under stage 4, a solid box states that the stake goes to the winner, or a public apology is made, and that losing artists face blacklisting and lost commissions. A dashed box below it lists exits after the vote: an intermediary planning to refund both sides on a near-tie, and a losing side refusing the result or payment.}
  \label{fig:bet-process}
\end{figure*}

Within Chinese online art commission communities (绘圈, \textit{huiquan}), there has been a rapid increase in the number of commissioners who believe that commissioned artists are using generative AI as part of their workflows without disclosing its use~\cite{Naojiti2026, Kazike2026}. Because artistic value in commission communities is often tied to perceptions of human effort, skill, and authenticity, allegations of undisclosed AI use can lead commissioners to feel misled and can undermine trust between artists and their clients~\cite{Sultana2026, hwang2026disclosure}. 

Commissioners pay a premium for work they believe to be human-made. However, research has shown that authorship labels alone can influence perceptions of an image's attractiveness,  profundity, or value, even when the image itself remains unchanged ~\cite{bellaiche2023humans}. Because commissioners often value effort and creative processes more than the finished product itself, they frequently rely on evidence such as work-in-progress recordings or provenance logs to judge whether a work is authentic ~\cite{Sultana2026}. If a final work is delivered without such evidence, it can be difficult for commissioners to quell concerns about its authenticity, especially in an era of widespread generative AI usage~\cite{lu2023seeing, roca2025detecting, Matatov2025}. Consequently, communities and commissioners have sought mechanisms that can generate process-based evidence of artistic skill and labor after a dispute arises~\cite{Naojiti2026, Kazike2026}. As a result, some art commission communities have begun leveraging a novel livestream-based adjudication practice, 对赌 (\textit{AI duidu}) to settle disputes about perceived generative AI usage. 

Three stakeholders are part of \textit{AI duidu}: the \textit{accuser} (i.e., the person who commissioned the artwork or a community member who suspects that AI was used in its creation), the \textit{accused artist} (i.e., the person who created the artwork), and an \textit{intermediary} (i.e., a third-party community member who oversees the process and holds any monetary stake in escrow). While the specific details of each \textit{AI duidu} may vary on a case-by-case basis, the process generally consists of four phases (Figure~\ref{fig:bet-process}). First, the accuser publicly challenges the artist to a \textit{duidu}. Second, if the accused artist accepts the challenge, both parties agree on the terms of livestream challenge, the designed intermediary, and the stakes involved (e.g., a public apology or a monetary payout two to three times the commission price). If a monetary stake is agreed up, the accuser and accused artist each provide their portion to the intermediary. Third, the accused artist livestreams themselves creating a new artwork whose concept and characteristics are designated by the accuser. The purpose of the livestream is to provide the community with insight into the artist's workflow and artistic skills. Fourth, community members who meet pre-specified eligibility criteria adjudicate whether the workflow and skills demonstrated during the livestream are consistent with the creation of the original disputed artwork. They then vote on whether they believe AI was used in the creation of the original disputed artwork.  Based on the voting outcome, the intermediary oversees the fulfillment of the agreed-upon terms (e.g., transferring the stake to the winner of the \textit{AI duidu}). 

The accused artist's livestream thus functions as \textit{community-generated provenance}, allowing observers to evaluate whether the artist possesses the skills and workflow necessary to have created the disputed artwork. At the same time, this mechanism is imperfect: a livestream cannot definitively prove whether AI was or was not used in a previous work, nor if it will be used in the future. The tension between protecting artists from accusations that may be difficult to disprove and protecting commissioners from undisclosed generative AI use makes the \textit{AI duidu} an especially interesting case study through which to examine how online communities negotiate and govern disputes over human creative labor. Our research thus contributes to the understanding of \textit{AI duidu} as a form of ``community-generated provenance'', whereby livestreams and wagering shift the verification of human authorship from the artwork itself to community evaluations of the artist's skill and creative process. 

\section{Research Context: Livestream-Based Dispute Resolutions in a Digital Art Community}
To better understand the \textit{AI duidu} phenomenon, we studied users within the community in which it emerged: a Chinese online art commission community (绘圈, \textit{huiquan}) centered on anime-style character illustration associated with \textit{erciyuan} (二次元) of anime, comics, games, and novels subculture in China~\cite{xu2025character, zhu2021shequan}. As a marketplace where artists and commissioners regularly negotiate artistic labor, reputation, and compensation, this community has been particularly affected by the growing availability of generative AI tools. This community thus offered a unique setting to study how members negotiated questions of authenticity, authorship, and trust in the age of generative AI.

\subsection{Community Members}
Commissioners in this community often had specific ideas or concepts they wished to realize and sought to leverage the skills of others, while some themselves were artists who collaborated with others to further develop their ideas. They commonly commissioned artwork for social occasions (e.g., birthdays) or personal uses (e.g., social media avatars), requesting depictions of fictional characters, fan art, or original characters they had created themselves. Because these requests were often highly personal, commissioners purchased more than artwork alone; they purchased \textit{the realization of an idea}. 

Members of this community participate across multiple social media platforms, including Weibo, QQ, and Xiaohongshu, where commissioners and artists arrange commissions directly, as well as dedicated commission platforms such as Mihuashi and Huajia~\cite{zhu2021shequan, Kazike2026}. Commission prices vary widely among artists, with those who have larger followings generally able to command higher rates, reflecting the strong economic incentives and reputational dynamics within the community ~\cite{zhu2021shequan, ma2026huajia}.

AI use was strongly discouraged within this community and community members expected artists to publicly oppose AI use. Although this community is aware of generative AI tools, they continued to commission human artists because they valued the labor, effort, and care invested in the creative process and viewed communication with artists as an integral part of the commissioning. For this community, the primary concern was not AI technology itself but the potential misrepresentation of the time and effort involved in hand-drawn labor if artwork was AI-generated. A secondary concern was that character designs, sketches, or concepts provided during the commission process could be input into AI models without the commissioner's permission. 

Although this community discouraged AI use, it did not share a single definition of what constituted AI use. The acceptability of generative AI for ideation, inspiration, image tracing, pose referencing, and automated brushes, among other practices, was often contested. More broadly, community members evaluated AI-assisted workflows based on whether human-skill, judgment, and creative agency remained central to the production process. 

\begin{table*}[t]
\centering
\footnotesize
\renewcommand{\arraystretch}{1.2}
\setlength{\tabcolsep}{4pt}
\caption{Overview of the 17 participants, including their occupation, age, role in the community, role in the \textit{AI duidu}, and the outcome of the \textit{AI duidu} they participated in.}
\label{tab:participants}
\begin{tabularx}{\textwidth}{@{}llcll>{\raggedright\arraybackslash}X@{}}
\toprule
\textbf{ID} & \textbf{Occupation} & \textbf{Age} & \textbf{\begin{tabular}[t]{@{}l@{}}Role in Commission\\Community\end{tabular}} & \textbf{Role in AI Duidu} & \textbf{AI Duidu Outcome} \\
\midrule

P1 & Former VTuber & 23 & Commissioner/Freelance Artist & Accuser & Canceled \\
P2 & Undergraduate & 18 & Commissioner & Accuser & Canceled \\
P3 & Undergraduate & 22 & Commissioner/Freelance Artist & Intermediary ($\times$2), Accuser ($\times$1) & Accuser Won ($\times$2), Canceled ($\times$1) \\
P4 & Undergraduate & 21 & Commissioner & Accuser & In progress \\
P5 & Undergraduate in Art & 20 & Freelance Artist & Intermediary & Canceled \\
P6 & Undergraduate & 18 & Commissioner/Freelance Artist & Accused Artist & Canceled; later reported cyber-bullying to the police \\
P7 & Game content operator & 24 & Commissioner/Freelance Artist & Accuser & Canceled \\
P8 & Unemployed & 32 & Freelance Artist & Accused Artist & Accuser Won \\
P9 & Undergraduate & 21 & Commissioner & Accuser & Canceled \\
P10 & Undergraduate & 18 & Commissioner/Freelance Artist & Intermediary & Canceled \\
P11 & Undergraduate & 18 & Commissioner/Freelance Artist & Intermediary & Accuser Won \\
P12 & Media operator & 26 & Commissioner/Freelance Artist & Accuser ($\times$1), Intermediary ($\times$1) & Accuser Won, In progress \\
P13 & New media operator & 26 & Freelance Artist & Accused Artist & Accuser Won \\
P14 & Undergraduate & 20 & Freelance Artist & Accuser & Accuser Won \\
P15 & Civil servant & 25 & Commissioner/Freelance Artist & Accuser & Accuser Won \\
P16 & Undergraduate in Law & 19 & Commissioner/Freelance Artist & Accuser & Accuser Won \\
P17 & Undergraduate in Education & 22 & Commissioner/Freelance Artist & Accuser & Accuser Won \\
\bottomrule
\end{tabularx}
\end{table*}

\subsection{Methodology}
Our exploration into \textit{AI duidus} was grounded in the experiences of those who had first-hand knowledge of them. Thus, we recruited participants through Xiaohongshu, a Chinese social media platform where many \textit{AI duidu} cases were publicly initiated, discussed, and voted on. We first identified Xiaohongshu posts related to \textit{AI duidu} and using the platform's direct messaging feature, contacted the users who had shared these posts. Our reach-out message is included in Appendix~\ref{appendix:reach-out}. We focused on users who had been involved in at least one \textit{AI duidu} as an accuser, an accused artist, or an intermediary. 

Seventeen members of this community agreed to participate in a semi-structured interview (Table \ref{tab:participants}). Participants ranged in age from 18 to 32 years (\textit{M} = 21.9, \textit{SD} = 3.8) and all identified as female. Most participants were both commissioners and freelance artists, although some occupied only one of these roles. Eleven participants had engaged with \textit{AI duidus} as accusers, five served as intermediaries, and three were artists who had been accused of using generative AI. Two participants (P3 and P12) had served as both accusers and intermediaries.

Before each interview, we obtained informed consent and reviewed our interview protocol (which was approved by our institution's Research Ethics Board), including the collection and use of data, participant rights, and confidentiality procedures. To protect participant's privacy, we removed or anonymized all identifying information, including online usernames and platform-specific identifies, from the study data and all reported findings. We conducted 16 interviews online, using WeChat. One participant declined voice communication and instead participated through text-based messaging. Interviews lasted around 60 minutes. Participants received an honorarium for their time, at \$16 CAD per hour. During the interviews, we asked participants about their perceptions of AI generated art, their experience participating in and watching \textit{AI duidus}, and their views on the role of livestreaming and wagering in verifying artistic authorship. 

After reading all 17 transcripts in the original Chinese and translating them into English, we identified three topics: why \textit{AI duidu} emerged and why participants took part in it, how it was carried out in practice, and how participants assessed its effectiveness, fairness, and consequences. Some of the questions that asked about these topics directly were: ``\textit{Why did you ultimately take the [AI duidu] route?}'' (accusers), ``\textit{Could you walk me through one [AI duidu] from beginning to end?}'' (intermediaries), and ``\textit{Do you think [AI duidu] can actually identify people who use AI, or is it more about setting rules and deterrence?}'' (accusers). However, answers to more general questions, such as those about participants' commission workflows and their views on AI, also contained discussions of these topics. Additionally, because all three protocols shared the same questions about the livestream and the community vote, we were able to compare how accusers, intermediaries, and accused artists described these stages. The author began by open coding two transcripts (P1 and P13), which produced an initial set of 49 codes, and then open coded all 17 transcripts, compared the new codes with this initial set, and arranged the codes into groups and hierarchies to determine emergent themes. Each code was given a definition with inclusion and exclusion criteria, and the resulting codebook was applied back to all 17 transcripts. When passages raised by at least two participants did not fit any existing code, a new code was added, which resulted in three additional codes. The final codebook contained 86 codes under six high-level themes: ``\textit{community ecology},'' ``\textit{commissioning},'' ``\textit{AI use and detection},'' ``\textit{AI duidu procedure},'' ``\textit{consequences},'' and ``\textit{reflections on the mechanism}.'' These themes broke down into more detailed codes. For example, ``\textit{AI duidu procedure}'' included ``\textit{finding an intermediary},'' ``\textit{negotiating the stake},'' ``\textit{voting mechanism},'' and ``\textit{withdrawal},'' while ``\textit{reflections on the mechanism}'' included ``\textit{whether AI duidu detects AI use}'' and ``\textit{refusal as an admission of guilt}.''

Our findings should be interpreted within the context of several limitations. First, the participant pool was relatively young, consisted entirely of women, and was drawn from a single Chinese online art commission community. While these characteristics reflect the demographics and social context of the community under study, they limit the transferability of our findings beyond this particular case. Second, our sample was skewed toward accusers and intermediaries, with comparatively fewer participants who had been accused of using generative AI. Finally, most participants described cases in which the accuser ultimately prevailed or the dispute was canceled, meaning that our analysis may underrepresent the experiences and perspectives associated with outcomes favoring accused artists. 

\subsection{Positionality Statement}
The research team recognizes the potential for our genders, cultural backgrounds, and social identities to shape our interpretation of the data. Within our team of four, two researchers identify as women and two as men, and three identify as White and one as Asian. The first author, an Asian woman and a native Chinese speaker, is familiar with the broader art commission culture and with platforms such as Xiaohongshu, which allowed her to conduct the interviews in Chinese and to interpret community-specific terms and practices in the original transcripts. The other three authors do not read Chinese and engaged with the data through the English translations.

\section{Related Work}
We first examine online art commission communities as marketplaces for personalized creative labor, focusing on how the value of commissioned artwork depends on commissioners' trust in the human effort, skill, and authenticity underlying the work, and how generative AI complicates this trust. We then review literature on the detection and perceived value of AI-generated art, drawing connections between people's limited ability to identify AI-generated images and the lower value they assign to works they believe were created using AI. Finally, we consider how online communities, including livestreaming communities, govern their members and resolve disputes through moderators, juries, and other forms of collective decision-making, to understand how communities may address disputes without relying on a permanent moderation authorities.

\subsection{Trust and Human Authorship in Art Commission Markets}

Online art commission communities operate as markets for personalized creative labor. Unlike many other creative markets, commissioners value not only the completed artwork but also the human effort, skill, and authenticity involved in its creation. Prior work on fan economies suggests that many commission communities inherited norms from gift-based creative cultures, where artistic labor carries social and moral value beyond market price \citep{hellekson2009fannish, turk2014fanwork, lessig2008remix}. Consequently artists frequently frame payment as compensation for labor and materials rather than profit-seeking activity \citep{kennedy2022fanbinding}. 

Research on craft economies and digital marketplaces suggests that consumers attach value to human handwork and authenticity \citep{luckman2013aura}. This premium depends on a buyer's ability to trust the human provenance of a product. In art commissions, value is similarly tied to beliefs about the artists' labor, skill, and creative process rather than solely to the completed artifact. Commissioners therefore purchase not only an image, but also the assurance that it reflects the artist's claimed skill sets. Evidence of process, such as work-in-progress images, sketches, or demonstrations of artistic skill, can help support such claims by making creative labor visible. However, because much of the production process remains private, commissioners, often have limited means of independently verifying how a work was created.

Generative AI challenges the ability of commissioners to evaluate whether creative work reflects the labor they are paying for. While prior research has documented disruptions to freelance labor markets and concerns about deskilling, style appropriation, and job displacement \citep{hui2024shortterm, helmsunderstanding, oztas2025reevaluating, shen2024soulmate, shen2021labor}, creative workers also face the new problem of credibility. Commissioners may wish to know whether AI was involved in production, while artists may struggle to communicate or verify the nature of their contribution. Existing disclosure and governance mechanisms remain inconsistent, and freelancers and commissioners often hold different expectations regarding AI transparency \citep{hwang2026disclosure, gao2026governance}. 

Existing studies have primarily examined the economic consequences of generative AI adoption or the experiences of individual creative workers. Much less is known about how creative communities collectively adjudicate disputes over suspected AI usage once trust has broken down, i.e., \textit{How do commissioners assess claims of human authorship? What forms of evidence are considered credible? How do communities balance protecting artists from unfounded accusations against protecting commissioners from deception? } We address these questions by examining how a commission-based community responds when the authenticity of its core product, human-created artwork, is called into question.

\subsection{Authorship Verification and AI Art Detection}

Understanding disputes over AI use first requires understanding whether people can reliably recognize AI-generated images. Existing evidence suggests that they often cannot. On a benchmark built from state-of-the-art generative models, human participants misclassified 38.7\% of images, even when the best detection model failed on 13\% of cases \citep{lu2023seeing}. Similarly, an online experiment with 12,500 participants worldwide, found overall detection accuracy was of only about 62\%, with substantial variation across image types \citep{roca2025detecting}. Additional large-scale evidence suggests that detection accuracy depends on factors such as scene complexity, artifact type, display duration, and whether AI images have been manually curated \citep{kamali2025photorealism}. These findings suggest that visual inspection alone cannot reliably establish whether a finished image was created by a human or generated using AI. As a result, communities concerned with undisclosed AI use may turn to process-based forms of evidence, such as demonstrations of artistic workflow and skill, when evaluating authorship claims. 

At the same time, believing that a work was AI-generated can reduce its perceived value. Experimental studies show that artworks labeled as AI-made receive lower ratings for aesthetics, composition, and expressiveness than otherwise comparable works labeled as human-made \citep{ragot2020aigenerated, hong2019artificial, bellaiche2023humans}. Human-attributed artworks also elicit stronger emotional responses \citep{agudo2022emotion, demmer2023emotional}. Importantly, \citet{grassini2024negative} showed that this bias is driven by perceived authorship rather than actual provenance. Although participants could not reliably distinguish human-created artworks from AI-generated images and collectively preferred the AI-generated images, they nevertheless devalued whichever works they believed were AI-made. Survey evidence points in the same direction: in one study, 79\% of respondents reported being unwilling to pay the same price for AI-generated art as for human-made work \citep{wlodarczyk2024digital}. These findings suggest that beliefs about authorship shape both the aesthetic and economic value of creative works, making accusations of AI use consequential for artists and commissioners alike.

As disputes over AI use are difficult to resolve through inspection of the finished artwork alone, some artists have incorporated time-lapses and livestreams into their community-building practices to demonstrate their end-to-end workflows and assure viewers or prospective clients of the authenticity of their work. On TikTok and Reddit, for example, artists have posted time-lapse videos of themselves recreating existing works or producing new artwork to counter accusations of using generative AI \cite{reddit_ai_accusation_2025, tiktok_ai_accusation_2026, tiktok_authorship_example}. Early work by Lu et al. demonstrated that livestreaming can support the preservation and transmission of cultural heritage by making specialized skills and creative labor visible to audiences \cite{lu2019streaming}. This visibility can, in turn, foster trust in the authenticity and value of traditional craft practices. Similarly, Fraser et al. interviewed artists and conducted content analyses of creative livestreams, finding that the transparency afforded by livestreaming provides viewers with an unedited view of otherwise invisible creative practices, helping build trust in artists' skills and expertise \cite{fraser2019sharing}. In follow-up work, Fraser et al. argued that while livestreams can illuminate artist's decision-making, mistakes, iterations, and work practices, the slow and labor-intensive nature of many creative processes can reduce viewer engagement and generate feelings of vulnerability among artists \cite{fraser2020livestreaming}. Consequently, many artists are reluctant to livestream their entire creative workflow. 

Taken together, these studies suggest that livestreaming can make creative expertise publicly observable. Consistent with prior work on reverse Turing tests, \citep{alizadeh2022reverse}, our findings show how livestreaming can function as a mechanism for authorship verification. Unlike prior forms of creative livestreaming that primarily support learning, community building, or audience engagement, \textit{AI duidu} emerges after accusations of AI use have already been made. Through a combination of livestreamed performance, community adjudication, and monetary wagering, verification shifts from the disputed artwork to the artist themselves. Rather than evaluating the artifact in isolation, community members assess the artist's skill, workflow, and ability to produce artwork in real time, transforming \textit{AI duidu's} livestreaming into a form of community-generated provenance.

\subsection{Online Community Regulation and Livestreaming}

Online communities regularly decide whether to warn, suspend, ban, or otherwise sanction members. The procedures for making these decisions vary, but a common model relies on one or more moderators who monitor behavior and rule on disputed incidents \cite{10.1145/3637296}. Moderators are often volunteers \cite{10.1145/3637296} who may be appointed by a founder or other moderators, or elected by the community. Prior work has shown that moderators collectively assemble evidence and review user histories to determine whether conduct violates community norms \cite{10.1145/3479554}. In making these decisions, moderators may consider factors such as intent, reputation, account age, expressions of remorse, and the risk of future harm. They often deliberate with other moderators or put cases to a vote. 

An alternative to delegating adjudication to a standing moderator is to distribute it among ordinary community members, most often through some form of jury. Fan and Zhang proposed digital juries as a civics-oriented approach to platform governance, in which randomly convened panels of users deliberate or blind votes on individual moderation cases \cite{10.1145/3313831.3376293}. They found that users perceived jury processes as more procedurally just than the platform governance practices they currently encounter. Subsequent work has examined whether such panels can bear the weight placed on them. Hu et al. reconvened the same jury to evaluate similar cases while concealing its awareness of prior deliberations, finding that juries were as consistent as individual decision-makers, partly because consensus emerged through the reinforcement of initial majority opinions \cite{10.1145/3411764.3445433}. Deployed systems illustrate both the promise and challenges of this approach. The League of Legends' Tribunal allowed players to adjudicate reports of disruptive behavior through non-hierarchical peer review \cite{10.1145/3134697}, while votekicking in social virtual reality extended collective sanctioning into synchronous, embodied settings, where decisions must be made in the moment and are difficult to appeal \cite{10.1145/3706598.3713577}. 

More recently, online communities have developed around the practice of livestreaming. Of particular relevance to this study, prior work has shown that livestreaming plays an important role in East Asian communities such as \textit{Otaku} communities \cite{10.1145/3411764.3445660}, shapes commerce and consumer behavior \cite{wang2022live}, and supports the development of fan communities both on and off livestreaming platforms \cite{lu2018you}. Prior research has also examined how moderators are appointed on Twitch \cite{10.1145/3290605.3300390}, how they manage toxic and anti-normative behavior, and how they impose sanctions \cite{10.1145/3452918.3458796}. In particular, Cai et al. found that livestreaming moderation involves highly visible and performative acts   \cite{10.1145/3452918.3458796} like declaring moderators' presence, reiterating community rules in chat, and providing real-time explanations for why certain behaviors are inappropriate. These public and communicative forms of moderation fostered engagement, strengthened community bonds, and educated members about community norms.  However, there remains a lack of work on how communities may self-govern or resolve disputes without replying on community\-appointed moderators. While the role of the intermediary discussed in this study shares some similarities with that of a moderator, intermediaries are appointed to adjudicate specific disputes rather than to enforce community standards more broadly. This work contributes one of the first studies examining how a community\-driven form of self-governance can be used to adjudicate disputes using livestreaming and wagering.

\section{Findings}
We first outline the \textit{AI duidu} procedure and the social dynamics that shaped each stage. We then examine the conditions that led to its creation and explore how participants assessed \textit{AI duidu's} effectiveness, fairness, and risks.

\subsection{AI Duidu Process}

An \textit{AI duidu} typically unfolds in four stages (Figure~\ref{fig:bet-process}): (1) community members voice their suspicion that an artist has used generative AI in a commissioned artwork, and either side proposes an \textit{AI duidu} as the dispute escalates, (2) both sides negotiate the rules and recruit an intermediary to hold their stakes in escrow, (3) the accused artist draws an artwork assigned by the accuser during a livestream so that viewers can watch the drawing process in real time, and (4) eligible artists and commissioners vote on whether the disputed artwork was created with AI, after which the intermediary announces the result and transfers the stakes to the winning side. The accuser or the accused artist can also withdraw at several points in this process (dashed boxes in Figure~\ref{fig:bet-process}), and many of the \textit{AI duidus} our participants described ended this way rather than through a formal verdict.

\subsubsection{Stage One: \textit{AI Duidu} Proposal and Accused Artist's Response}

\begin{figure*}[t]
  \centering
  \includegraphics[width=\textwidth]{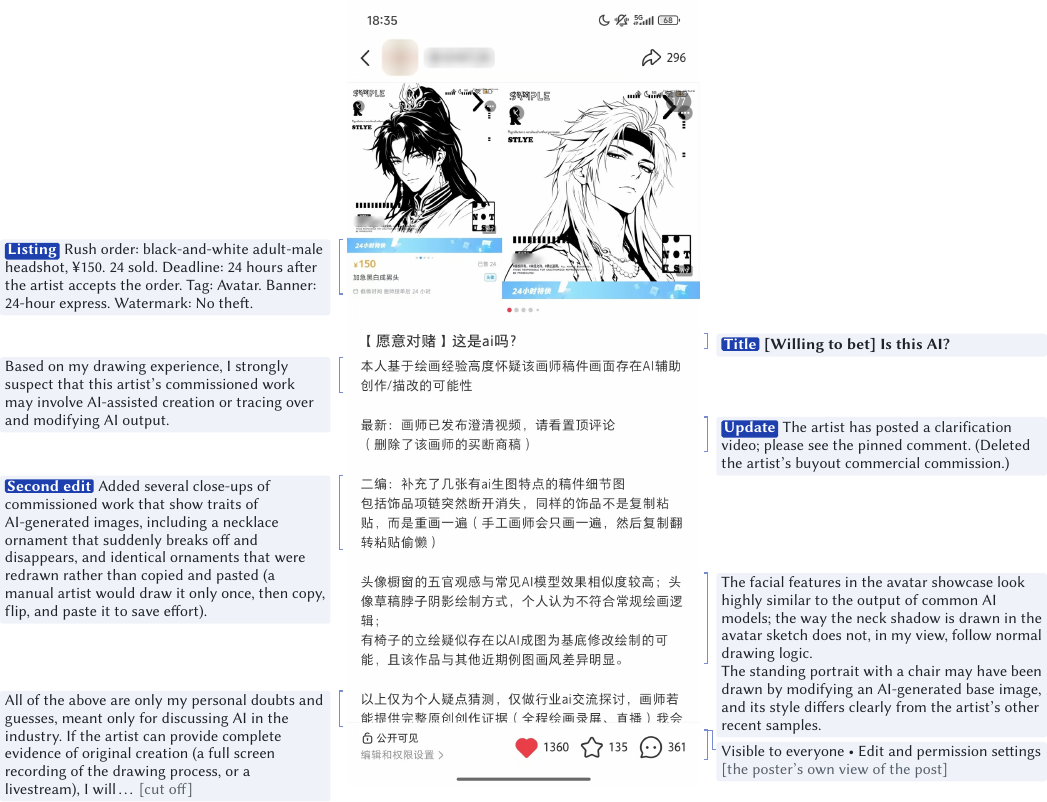}
  \caption{The public Xiaohongshu post by P7 (Accuser, Artist/Commissioner) accusing an artist of AI use, with our English translation (Brackets indicate the Chinese text that each box translates). The post lists the cues P7 relied on, including a necklace that breaks off, ornaments redrawn instead of copied, AI-like facial features, and a style that differs from the artist's other samples. Its title offers to settle the question with an \textit{AI duidu}. The poster's avatar and handle, and the accused artist's handle, have been intentionally blurred.}
  \Description{A phone screenshot of a Xiaohongshu post, with English translation boxes on both sides of it, each level with the text it translates. At the top are two black-and-white character illustrations from the accused artist's shop, one listed as a 150 RMB rush black-and-white male headshot. Below them, the post, titled ``[Willing to bet] Is this AI?'', explains why the poster suspects AI-assisted creation. It points to a necklace that breaks off, redrawn ornaments, AI-like facial features, unusual neck shading, and a style unlike the artist's other samples, and asks the artist for a full screen recording or a livestream.}
  \label{fig:p7-post}
\end{figure*}

All \textit{AI duidu} began with suspicion from unsolicited exposure to disputed work. Five accusers (P1, P12, P14, P16, P17) first encountered such works through algorithmic feeds. P14 recalled that \textit{``the AI traces were just too obvious''} and added that in a small fandom `\textit{`our circle is too small, so it was very easy to come across them.''} Other participants encountered suspicious works while browsing commission platforms with automatic tagging (P4, P7), after receiving their own commissions (P2), or through images forwarded by friends (P14). Platforms sometimes raised the initial warning. In P12's case, Mihuashi identified images as AI-generated, after which the artist protested publicly. P8 first learned of the suspicion through a risk warning placed on their shop page. Participants also encountered suspected works in certified pools and featured rankings (P2, P10). Such cases weakened confidence in platform vetting systems.

Participants drew on several types of cues when suspicious arose. Local structural anomalies such as warped lashes, fused color blocks, extra fingers, contradictory lighting, and mismatched ears were frequently cited (P4, P5, P9, P11, P15). P7, who listed such cues in a public post (Figure~\ref{fig:p7-post}), explained that ``\textit{the braided chain simply breaks off where it meets the bead chain \ldots a manual artist would never produce this, it is classic AI generation, it swallows exactly these small details.}'' Other cues included uneven levels of finish within a single image (P2, P3, P15), a perceived ``\textit{AI flavor},'' recognition of specific generative models (P3, P16, P17), and inconsistencies with an artist's earlier work (P3, P7, P9, P15, P16). Participants also cited process-based evidence. For example, P2 noted cases in which a highly refined element appeared in a layer file despite having no visible underpainting. P14 described one artist as having ``\textit{copied even the errors}'' of a generated image. At the same time, participants  rejected some popular heuristics such as Photoshop threshold tests, and acknowledged that some human-made styles were incorrectly flagged because they resembled AI-generated images.

Once participants became suspicious, they either remained silent or began voicing their concerns. Silence was the default response. P7 stated that artists ``\textit{generally do not speak up about artists I merely have some doubts about}''. P8 described a similar tendency. However, when someone eventually voiced a concern, others often responded that ``\textit{finally someone said it, some people had been holding it in for a long time, afraid of the artist's fans}'' (P14). Participants often raised concerns in private channels with friends (P2, P14) or in QQ groups before moving discussions to Xiaohongshu (P7). Disputes over whether an artist had used AI frequently escalated through comments and direct messages, where exchanges sometimes deteriorated into insults and personal attacks (P2, P4, P8, P12, P17). As suspicions accumulated and became more widely shared, participants increasingly sought ways to either substantiate or refute the allegations.

The accused artist's response strongly shaped whether participants' suspicions escalated into \textit{AI duidu}. Defiant public replies were a common source of escalation. P4 proposed a \textit{AI duidu} after receiving a response in an ``\textit{unfriendly tone}.'' P14 described an artist who ``\textit{did not admit anything, and many people went to attack those who questioned him \ldots he was too brazen, and that genuinely enraged me.}'' In private channels, artists sometimes demanded that posts be deleted (P16), sent insults (P8), or proposed a \textit{AI duidu} themselves (P9, P13).

When disputes escalated, either side could propose \textit{AI duidu}, and accused artists did so surprisingly often. Some responded directly to warning posts by saying ``\textit{stop accusing me, why not sign a bet contract with me}'' (P1). Others proposed a \textit{AI duidu} through direct messages (P9, P13) or public comments that specified a stake (P8, P12, P14). When accusers initiated a \textit{AI duidu}, they often described it as a response to perceived stalling (P2), a hostile tone (P4), or an impulsive decision made in the heat of a dispute (P16).

Attempts at self-vindication could also intensify suspicion. In several cases, participants interpreted the artist's attempts to prove their innocence as additional evidence against them. P15 described a case they later pursued: \textit{``He posted a self-proof video saying that every accusation made him physically sick, and in the video he produced a picture with almost no changes at the drafting stage''}. P7 said that one artist's proof video ``\textit{did not even include a draft, it started straight from line work, which does not match any normal drawing process.}''

\subsubsection{Stage Two: Rules and Pre-Livestream Setup}

\begin{figure*}[t]
  \centering
  \includegraphics[width=\textwidth]{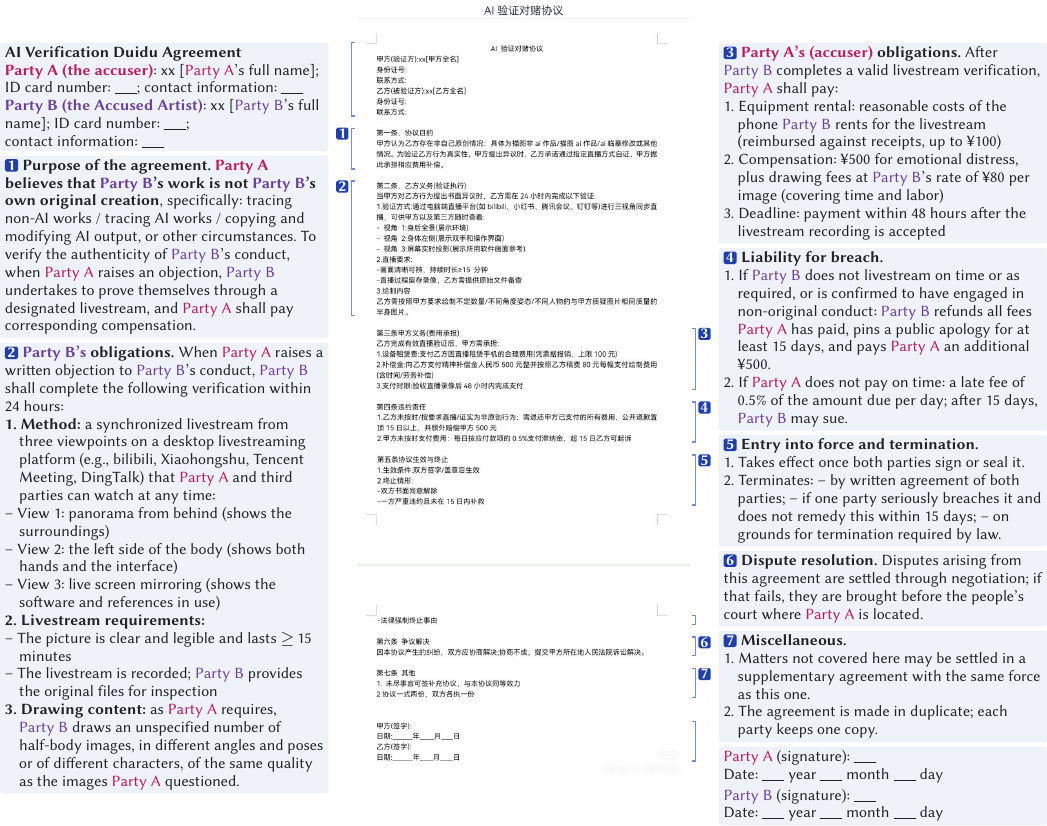}
  \caption{A written \textit{AI duidu} agreement template shared by P7 (Accuser, Artist/Commissioner), with the English translation. She sent the agreement to the accused artist to sign in Stage 2: Rules and Pre-Livestream Setup.}
  \Description{A two-page Chinese document titled ``AI Verification Duidu Agreement'', with its English translation in two columns, one on each side of it. Party A, the accuser, and Party B, the Accused Artist, each have blank fields for a name, an ID card number, and contact information. Seven articles cover the purpose of the agreement, Party B's duty to livestream from three viewpoints within 24 hours, Party A's costs and compensation, liability for breach, entry into force and termination, dispute resolution in the people's court, and miscellaneous terms, followed by signature and date lines.}
  \label{fig:agreement}
\end{figure*}

Once an \textit{AI duidu} was proposed, both sides faced pressure to participate. For the accused artist, refusal could be interpreted as evidence of guilt. P1 noted that refusing an \textit{AI duidu} could appear suspicious but cautioned that ``\textit{this cannot be decisive evidence, there are plenty of socially anxious artists.}'' Accusers also faced pressure. As P15 explained, \textit{``if you claim a work is AI and you do not bet, your words are not credible, you are defaming the author, and only through the bet can you prove your own legitimacy."}

Rules governing an \textit{AI duidu} were typically negotiated on a case-by-case basis, although community rule templates sometimes existed (Figure~\ref{fig:agreement}). Some \textit{AI duidus} had no written agreement at all (P5, P8, P12, P14, P15), while others specified only voting procedures. Rules could also change over time. Participants described modifications occurring the day before a livestream (P14), during the livestream itself (P13), or even after the result was announced. P8 recalled learning only after losing that the rule had never been \textit{``I finish the drawing, prove it is handmade ,and I win.''} Attempts to formalize the process were rare.  When P7 researched local notarization offices and shared the associated costs, the accused artist  withdrew from the \textit{AI duidu}. 

An intermediary was recruited to oversee the process and hold the stake. Although participants likened intermediaries to  ``\textit{a piggy bank}'' (P5), ``\textit{basically a Xianyu escrow platform}'' (P7), and ``\textit{an impartial machine}'' (P1), their responsibilities extended far beyond escrow. Intermediaries watched streams, captured evidence, prepared and checked voting surveys, published announcements and results, counted votes, mediated rule disputes, moderated Danmaku, and sometimes supported both parties emotionally. Recruitment was based largely on reputation, prior experience, artistic expertise, perceived neutrality, and account credibility (P1, P2, P5, P7, P12, P14, P15, P16). Nevertheless, some participants conducted little vetting because they viewed the intermediary primarily as a witness rather than a guarantor (P2, P8, P17).

Stakes were commonly set at two to three times the accused artist's commission price (P1, P6, P10, P14, P15, P16), although some \textit{AI duidu} used unrelated amounts (P9, P11) or non-monetary penalties such as public apologies (P13). Both sides used  stakes strategically. Artists sometimes proposed high stakes to discourage challengers (P12, P15), while accusers lowered  stakes to encourage participation. As P5 explained, ``\textit{the money is not the point, the point is that everyone knows.}''  Participants viewed escrow primarily as a mechanism for discouraging withdrawal and compensating falsely accused artists, though noted that \textit{``when the stake has actually become payable is very hard to determine.}'' 

Once the rules, intermediary, and stakes had been established, participants negotiated the livestream requirements. The livestream assignment was the accuser's primary tool for testing the artist. A complete artwork was usually required (P9, P10, P14, P16), although some participants believed a finished detail or even a sketch could be sufficient (P2, P7). In seven cases, the accuser selected the topic. In these cases, the topic was revealed only after the livestream began. P3 described this practice as ``\textit{leaving no time at all to use AI.}'' In other cases, artists were allowed to choose one image from a set prepared by the accuser (P12, P14, P16). Assignments were typically designed to resemble the disputed artwork in character, angle, or style so that livestream viewers could directly compare the two works (P3, P5, P7, P9, P14, P16, P17). Participants also negotiated camera placement, workstation visibility, and screen-sharing requirements. These negotiations often exposed tensions between transparency and privacy. P13, an accused artist, for example, refused to show their hands and surroundings on camera because they viewed the requirement as degrading. 

\subsubsection{Stage Three: The Livestream}

\begin{figure*}[tp]
  \centering
  \includegraphics[width=\textwidth]{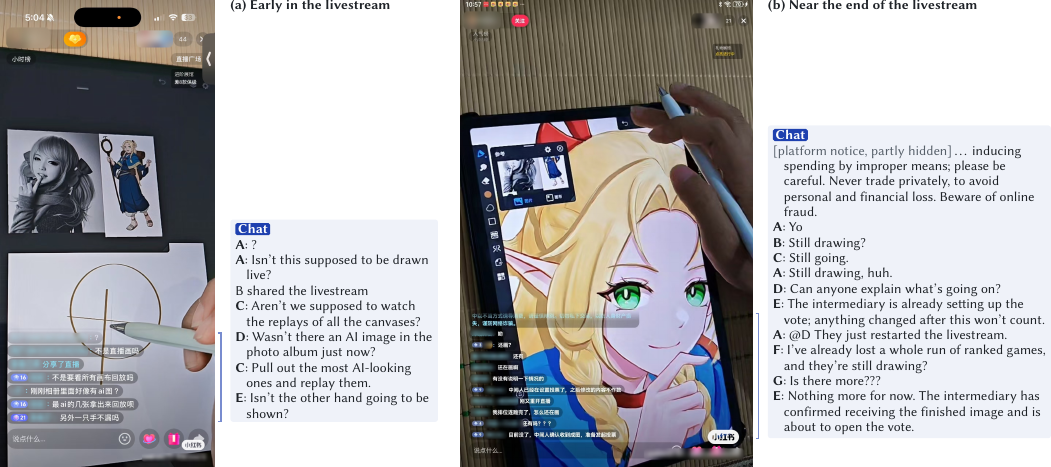}
  \caption{Two moments from the livestream of an \textit{AI duidu} that P11 (Intermediary, Artist/Commissioner) oversaw, with the live chat translated into English (letters are used in place of viewer names within each screenshot). (a) Early in the stream, a camera films the tablet, the artist's hand, and two reference images. Viewers asked for canvas replays, asked whether an AI image had just appeared in the artist's photo album, and asked to see the other hand. (b) At the end, viewers learned that the intermediary received the finished image and was setting up the vote. Handles, avatars, viewer names, and Xiaohongshu IDs are intentionally blurred.}
  \Description{Two screenshots of a Xiaohongshu livestream, each with its chat translated beside it. In the first, a camera looks down at a drawing tablet showing a grayscale photo and an anime character as references, while the artist's hand holds a stylus over a circle guide. Viewers ask whether this is supposed to be drawn live, ask to see replays of all canvases, ask whether there was an AI image in the photo album, and ask to see the other hand. In the second, the camera shows a nearly finished colored drawing of the character on the tablet. A viewer explains that the intermediary has confirmed receiving the finished image and is about to open the vote, and that changes made after this will not count.}
  \label{fig:livestream-p11}

  \bigskip
  \includegraphics{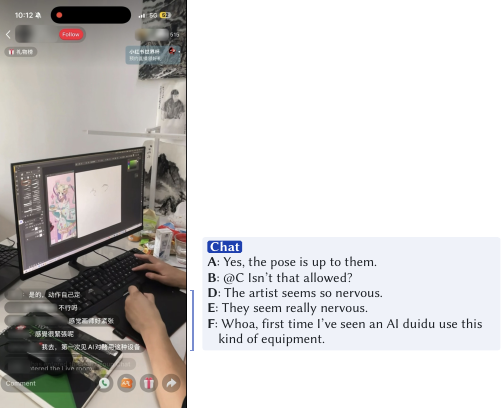}
  \caption{Another \textit{AI duidu} live stream, with the live chat translated into English (letters are used in place of viewer names). A phone camera filmed the accused artist's desk, monitor, keyboard, and pen tablet. Viewers remarked on how nervous the artist seemed. Handles, avatars, and viewer names are intentionally blurred.}
  \Description{A screenshot of a Xiaohongshu livestream filmed from behind the artist's shoulder: a monitor shows drawing software with a colored character illustration beside a nearly blank canvas, and the artist's hands rest on a keyboard and a pen tablet. The translated chat reads: yes, the pose is up to them; isn't that allowed; the artist seems so nervous; they seem really nervous; first time I've seen an AI duidu use this kind of equipment.}
  \label{fig:livestream-desk}
\end{figure*}

The livestream usually took place on the platform where the dispute had developed (P1, P2, P3, P4, P11). It often used the accused artist's established account, which some participants treated as an additional form of collateral (P4). The goal was to make the drawing process continuously observable, so screen mirroring alone was considered sufficient. P3 explained that the objective was ``\textit{to see you draw stroke by stroke, truly by yourself, not with an image lying beside you that has already been generated}.'' A stricter setup used two devices, with one streaming the screen and another filming the artists' hands, tablet, and desk (P11; Figure~\ref{fig:livestream-desk}). Some cases also limited screen brightness because an overexposed camera view could conceal a reference image beneath the drawing (P15). Continuous observation was itself treated as a safeguard because viewers could notice if a prepared image was introduced during the stream (P11; Figure~\ref{fig:livestream-p11}).

Livestream audiences varied from as few as three or four viewers (P8) to five or six hundred (P13, P17), and occasionally thousands (P11). Audience size depended on recommendation traffic and the artist's existing visibility (P1, P2). Stream length ranged from less than two hours to overnight. Participants described two hours as the norm, adopted in part to make the process more humane after earlier cases became excessive (P7). Artists also differed in how they interacted with viewers. Some chatted, took song requests, or discussed equipment (P15, P17). Others remained silent (P8, P16) or closed the Danmaku because they found it difficult to tolerate (P13). 

Accusers and intermediaries also remained active throughout the stream. Accusers monitored the drawing process, recorded footage, moderated comments, and sometimes faced abuse before any verdict existed (P15). Intermediaries moderated the room, captured evidence, and often withheld their own opinions to appear impartial (P5, P11). Accusers and intermediaries sometimes removed hints from the chat (P11, P15) or muted dissenting viewers (P17). 

\subsubsection{Stage Four: Community Vote and Outcome}

\begin{figure*}[t]
  \centering
  \includegraphics[width=\textwidth]{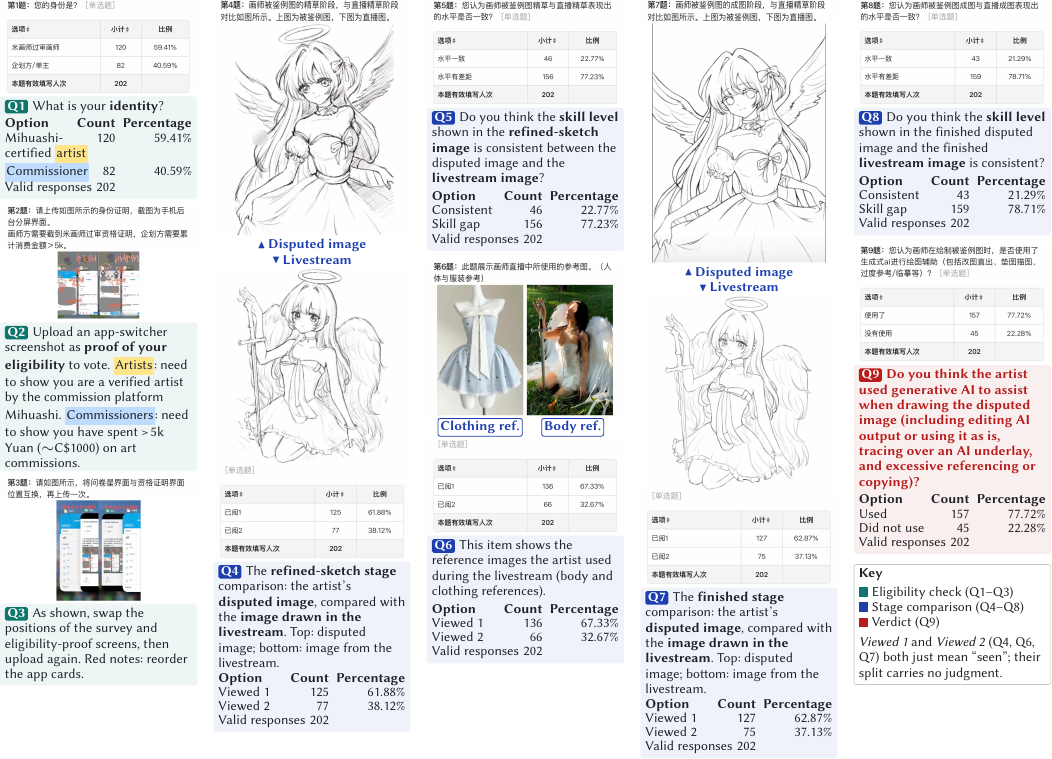}
  \caption{The voting survey posted by P15 (Accuser, Artist/Commissioner), with our English translation of each question and the results tables for each question. Voters first proved their eligibility (Q1-Q3), then compared the disputed image with the livestreamed drawing at the refined-sketch and finished stages (Q4-Q8), and finally judged whether the artist used generative AI (Q9). In this case, 157 of 202 voters (77.7\%) judged that the artist had used it. The artist's signature (Q4) and an account ID (Q2) were intentionally blurred.}
  \Description{Results pages of a Chinese online survey posted by P15, arranged in five columns. Under each question, an English box translates the question and its results table. Questions 1 to 3 check whether the voter is a Mihuashi-certified artist or a commissioner who has spent more than 5,000 RMB, using app screenshots as proof. Questions 4 to 8 place the disputed image above the livestreamed drawing at the refined-sketch and finished stages, show the reference photos the artist used, and ask whether the skill levels match; most voters answered that they do not. Question 9 asks whether the artist used generative AI; 157 of 202 answered yes.}
  \label{fig:survey}
\end{figure*}

The voting process used online surveys that asked for proof of the voter's eligibility, questions about the content in the livestream at various points, and if the viewer thought that the artist had used generative AI (P11, P16; Figure~\ref{fig:survey}). Voters typically compared the disputed work, process images from the livestream, and the final livestreamed artwork. Some cases also shared stage-by-stage comparisons or accelerated recordings of the livestream for those who missed the stream (P7, P14, P16). 

Voting was generally restricted to recognized artists or experienced commissioners. As P14 explained, ``\textit{an artist can see the technique, and a commissioner who has seen enough commissions can see the clues.}'' Passing Mihuashi's review process and becoming a certified artist on the platform often served as an eligibility criterion, with P15 describing it as ``\textit{the civil-service exam of artists.}'' As commissioners, voters were often required to demonstrate prior spending on art commission by showing payment records; as artists, they needed certification on a commission platform (such as Mihuashi), minimum commission prices, follower counts, or some combination of these criteria (P5, P11, P12, P13, P14, P15, P16, P17). Intermediaries sometimes adjusted these thresholds in response to community feedback or for more difficult cases (P15, P17). Participants nevertheless acknowledged that such eligibility criteria were imperfect:  some high-spending commissioners could not identify AI use (P15), while platform certifications did not guarantee AI-detection expertise (P16, P17). 

While voting, participants focused on drawing technique rather than surface resemblance. As P1 explained, voters assessed ``\textit{whether the technique is the same, not merely whether the images look alike}.'' They examined whether unusual errors from the disputed image reappeared during the livestream, with P14 describing one artist who ``\textit{reproduced the classic error perfectly, he copied even the mistakes.}'' Participants also considered line quality, drafting behavior, time spent on difficult details, signs of tracing, recurring habits when drawing faces and hair, body proportions, composition patterns, and inconsistencies between different artistic skills (P5, P11, P12, P15). Experienced voters paid particular attention to color-selection logic and grayscale value structure, which P11 argued many casual viewers could not evaluate. Ultimately participants described judgment as a form of trained intuition developed through repeated exposure rather than a formal detection method (P11, P13, P16). 

\begin{figure}[t]
  \centering
  \includegraphics{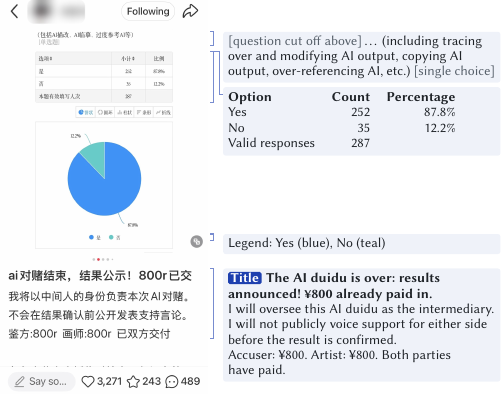}
  \caption{P3 (Intermediary/Accuser, Artist/Commissioner), as the intermediary, made a Xiaohongshu post announcing the result of an \textit{AI duidu}, shown with our English translation. P3 promised not to voice support for either side before the result was confirmed and stated that the accuser and the artist have each paid 800 RMB. Of 287 respondents, 252 (87.8\%) answered yes to the question of whether the artist used AI, resulting in the accuser's win. The poster's avatar and handle are intentionally blurred.}
  \Description{A screenshot of a Xiaohongshu post by P3, the intermediary, with English translation boxes beside it. A survey results table and pie chart show 252 yes answers (87.8 percent) and 35 no answers (12.2 percent) out of 287 to a question about AI tracing, AI copying, or over-referencing AI. The post is titled ``The AI duidu is over: results announced! 800 RMB already paid in'', and its author states that they oversee the AI duidu as the intermediary, will not publicly support either side before the result is confirmed, and hold 800 RMB from each party.}
  \label{fig:vote-result}
\end{figure}

Accusers won most bets. P2 estimated that successful self-vindications numbered only ``\textit{maybe ten or twenty}'' in total, while P3 attributed this imbalance to a selection effect, arguing that artists without visible problems were less likely to be challenged. Voting outcomes were not always decisive, however. Some cases gave greater weight to artist votes (P15, P17), while in others the opinion of a prominent artist effectively settled a dispute that community voting could not resolve (P5, P7, P13, P14, P15). Once a decision was reached, the intermediary typically transferred the stake to the prevailing side (P1, P3, P11, P16) or enforced a public apology. Intermediaries also published the results of the vote online (Figure~\ref{fig:vote-result}).

Participants largely viewed money as secondary to the reputational judgment produced by the process. For losing artists, the primary consequences were social rather than material, including blacklisting, public warnings, and loss of commissions (P3, P10, P11, P13, P17) and, though mass reporting, the removal of platform certifications (P9, P11, P16). Others accepted the result and issued partial refunds to commissioners (P16). These sanctions were limited because, while the community could warn others, it lacked the capacity to impose durable punishments. As P13 observed, with account-based sanctions, ``\textit{if he can open one new account he can open five}.'' Some artists deleted their accounts (P3), hid posts and refunded earlier commissioners (P15), or continued to deny the accusations (P14, P15), particularly when an artist's customers were outside the community discussing the \textit{AI duidu} (P8). 

Taken together, these outcomes suggest that the principal effect of \textit{AI duidu} was reputational rather than institutional. The process could shape community perceptions of artists and accusers alike, but it provided few mechanisms for enforcing long-term sanctions or compensation.

\subsubsection{Withdrawing from an \textit{AI Duidu}}
Many \textit{AI duidu} ended not through a formal verdict but through the withdrawal of one of the participants. Withdrawal could occur at several points in the process. Some negotiations collapsed before the terms were finalized: P1's counterpart abandoned a drafter contract, and one artist left after P7 challenged their legalistic bluff. Other \textit{AI duidu} stalled when a stake was never deposited (P9, P10), when a party disappeared before the scheduled stream (P2), or after a completed vote when the losing side refused the result or payment (P12, P15).

Accused artists withdrew most often, although accusers sometimes withdrew as well because they risked public criticism if the artist was ultimately cleared (P11). Intermediaries could also withdraw. In one case, an intermediary absconded with the funds (P6); in another, the intermediary was going to refund both sides rather than rule on a nearly tied vote until a prominent artist intervened (P1). The reputational meaning of withdrawal also depended on who exited. When an accused artist withdrew, participants often interpreted the act as indirect confirmation of AI use. In contrast, an accuser's withdrawal was more likely to be treated as an admission of defeat. 

\subsection{The Origin of the \textit{AI Duidu}}

Within this community, the \textit{AI duidu} emerged in response to three interrelated concerns. First, commission platforms were perceived as unable to adjudicate suspected AI use. P2 explained that platform reviews typically examined a finished image and a limited number of process images rather than the complete drawing process. P2 recalled a prominent Mihuashi artist who appeared in a featured ranking and was later accused of AI use. For P2, the case demonstrated that platforms struggle to identify AI involvement once generated material has been sufficiently modified to appear less obviously synthetic.
    
Second, participants described having strong emotional desire for redress. P8, despite having been accused of AI use, understood why commissioners felt like they needed to obtain both emotional and evidentiary closure: \textit{``If the artist does not take it seriously, feeds your commission to AI and sends it to me, I cannot tell, I am not someone who draws. I accept the piece, I am happy, I post it online, and then people say it is AI. I feel my feelings and my money have been hurt, so I hate artists who use AI to produce my image. But I do not understand it myself, so I need an outlet to prove whether this artist used AI. Call it an emotional outlet or an outlet for proof, and so this practice exists.''}

Third, participants repeatedly described generative AI as destabilizing the trust on which commission transactions depend. At the same time, they saw no authoritative method for restoring that trust. At the same time, they saw no authoritative means of restoring trust. P11 explained: \textit{``Once AI drawing became widespread, the entire trust chain from artist to commissioner simply snapped, because there is no official tool or platform feature that can determine one hundred percent whether a finished piece involved AI so such a suspicion has no resolution.'' }
    
Against this backdrop, livestreaming emerged as a viable mechanism for adjudication. Because the assignment was revealed only after the stream began and the drawing process was collectively witnessed in real time, livestreams limit opportunities to prepare AI-generated material in advance. Before the \textit{AI duidu} became established, participants believed that even process recordings could be fabricated with AI. P5 explained: \textit{``Before the AI duidu, some artists under suspicion would release the drawing process of the disputed image. But because a drawing process can now also be faked with AI, people instead watch carefully during the livestream, at how the artist sets down the draft, how they color, and whether the proportions can reach roughly the level of that earlier piece.'' }

The earliest version of the \textit{AI duidu} involved no monetary stakes. P5 recalled that money was added later: \textit{``At the very beginning, in my memory, the bet involved no money at all, there were not even the notions of accuser and accused. A group of people simply felt that an artist was suspected of using AI and asked them to open a livestream to prove themselves. But later everyone discovered that without money, first, the artist might simply run off, and second, if the artist had really drawn it themselves and had been wrongly accused, the money would serve as compensation to the artist. So gradually it evolved into a bet that requires stakes.'' } The introduction of stakes raised the cost of withdrawal for the accused artist, while also providing compensation for artists who were publicly accused but ultimately cleared of wrongdoing. 
Introducing money, however, created another problem: neither side trusted the other to hold the stake. As P5 explained, the intermediary emerged to safeguard the funds and support the perceived fairness of the process.

\subsection{\textit{AI Duidu's} Perceived Effectiveness, Limits, and Risks}
As \textit{AI duidu's} were privately organized, lacked legal protection, and depended primarily on public scrutiny, rule-setting power often belonged to the people who already understood community conventions. This could disadvantage outsiders. P8, an accused artist, for example, entered a bet believing that completing a drawing during a livestream would be sufficient. They later described the process as ``\textit{whatever rules, all that mess, it was just whatever they said, I did not understand, so whatever they said counted.}''  P12 similarly described an opponent who did not understand the community, basic artistic conventions, or the process itself. People who were less embedded in the community could therefore be both more likely to be accused and less prepared to negotiate the rules used to judge them.

Participants believed that \textit{AI duidus} worked best when the accused artist lacked the skill needed to produce work comparable to the disputed image. P3 stated that ``\textit{when your level clearly cannot keep up with AI, you are filtered out very easily}.'' P7 argued that ten minutes of sketching could be sufficient in obvious cases. The test became much less decisive when the accused artist was already highly skilled. P5 described how \textit{``There was a famous case where the artist genuinely had the ability to draw the disputed image, and his past works matched it. Many people still believed the image had been retouched with AI. But if an artist can really produce work matching his AI-generated image, it is very, very hard to judge through a livestream whether it is real or not.''} 

As a result of the livestream format and rules, many accused artists felt that the livestreams imposed restrictions on ordinary work habits and bodily movement. One case limited bathroom breaks to five minutes and allowed only snacks on the desk (P16). P13 noted that the format disrupted normal habits such as standing, resting, and consulting references. P7 similarly observed that artists sometimes avoided ordinary corrections because they feared that revising a line would appear suspicious. P13, an accused artist, who drew for four hours after a full workday, reported that early mockery damaged their concentration and resulted in an artwork they considered only ``\textit{somewhat unsatisfactory}.'' In this way, the livestream not only served as a mechanism of verification but also imposed significant psychological and behavioral constraints on the accused artist.

Other failure modes bypassed the test rather than challenging its interpretation. Participants considered the use of a hired stand-in difficult to detect in principle (P2). This risk became concrete in P12's case when the person negotiating through the account turned out to be a spokesperson for another artist who lived near an art academy. Off-camera references also created another loophole. P12 acknowledged that livestreams' relaxed camera setups could not rule out these risks. P15 identified a different problem: a skilled artist may have used AI simply to save time and effort. If that artist can successfully recreate the image when challenged, the livestream cannot establish whether AI was used in the original artwork.

Participants acknowledged several ways in which intermediaries could fail. Although intermediary were expected to support fairness, they possessed neither formal qualifications nor institutional protections. P11 stated that ``\textit{intermediaries like us have no legal qualification whatsoever, there is no guarantee behind our responsibility, and partiality, absconding, and fund disputes all happen easily}.'' Some intermediaries were also described as plants (P5), accepted bribes or tip-offs (P8, P11), received percentage kickbacks (P4), collected evidence on behalf of accusers (P4), or adopted unnecessarily hostile tones when questioning artists during livestreams (P13). 

\begin{figure*}[t]
  \centering
\includegraphics[width=\textwidth]{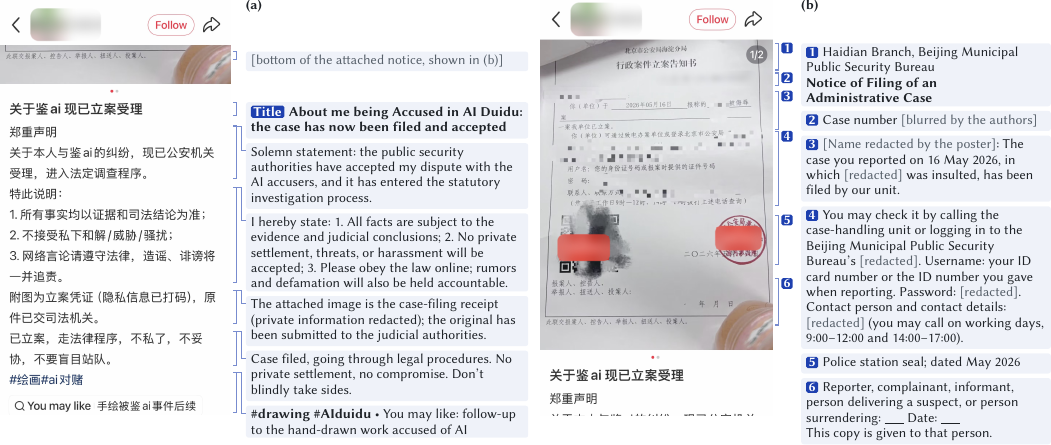}
  \caption{After being required to participate in an \textit{AI duidu} as the accused artist and subsequently being harassed online, P6 (Accused, Artist/Commissioner) reported the incident to the police and posted online that the police had accepted the report (a), attaching a notice stating that an administrative case for insult had been filed (b). The poster's avatar and handle, and the case number are intentionally blurred. The other redactions are the poster's own.}
  \Description{Two screenshots of one Xiaohongshu post, each with an English translation beside it. The first shows the post text, titled ``About me being Accused in AI duidu: the case has now been filed and accepted''. The poster states that the police have accepted their dispute with the AI accusers, that they will not accept private settlement, threats, or harassment, and that rumors and defamation will be held accountable. The second shows the attached notice from the Haidian Branch of the Beijing Municipal Public Security Bureau, stating that the case reported on 16 May 2026, in which a redacted person was insulted, has been filed.}
  \label{fig:p6-police}
\end{figure*}

Participants emphasized that reputational judgments could carry significant personal consequences including public ridicule, account abandonment, and have negative effects on mental health (P3, P6). P6, an accused artist, later reported the cyber-bullying to the police, who opened an administrative case (Figure~\ref{fig:p6-police}). Reflecting on these outcomes, P6 remarked that ``\textit{people announce who won, but few ever speak of the psychological problems.}'', while P7 warned that a community built around public accusations, particularly one involving many underage members, risks causing serious harm when disputes escalate into personal crises. 

Despite these limitations, most participants still saw value in the process because they understood its primary function as deterrence rather than definitive verification. P3 rejected the idea that a livestream could settle a case ``\textit{once and for all}'' and instead described it as ``\textit{a filter.}'' P9 similarly characterized the mechanism as a way of ``\textit{setting rules}.'' P12 argued that public scrutiny raises the cost of quietly using AI and therefore ``\textit{does more good than harm.}'' 

Participants accepted the mechanism because they saw few workable alternatives for repairing trust after an accusation. They nevertheless proposed several ways to address its weaknesses. Before \textit{AI duidu} became common, suspected artists often livestream informally, shared layer files, posted process videos, or responded to accusations with annotated explanations (P6, P15). Participants believed these practices gradually lost credibility as they became easier to fabricate. Suggested improvements included using an e-commerce platform such as Xianyu to hold funds and arbitrate payments (P14), observing artists in person to eliminate off-camera loopholes (P16), replacing human intermediaries with software or bots (P11), or building a dedicated platform with certified voters (P14).

\section{Discussion}

\subsection{Distinguishing Human-Created and AI-Generated Art}
A central tension underlying \textit{AI duidu} is that artists and commissioners often care deeply about whether an artwork was created by a human, yet neither humans nor existing detection systems can reliably distinguish AI-generated artwork from human-created work. Rather than producing definitive evidence, community members often relied on indirect signals such as artistic history or stylistic consistency when evaluating authorship claims. 

This uncertainty places an additional burden on artists. Participants described maintaining archives of sketches, project files, and recordings of their creative process to demonstrate that their work was human-made. Such expectations suggest that artists are increasingly required to authenticate not only their artwork but also themselves as creators. Moreover, because sudden stylistic changes or experimentation could attract suspicion, artists may feel pressure to maintain a consistent artistic identity rather than explore new creative directions. In this sense, uncertainty surrounding AI authorship can constrain creative expression by making stylistic experimentation socially risky.

More fundamentally, these findings raise questions about why human authorship matters. When commissioners express a preference for ``human-made'' artwork, it remains unclear whether they are responding to perceived differences in artistic quality \cite{bellaiche2023humans, ragot2020aigenerated, hong2019artificial}, valuing the human labor embodied in the work \cite{ lu2019streaming}, or adhering to community norms \cite{seering2019moderator, gieryn1983boundary} that define human authorship as the more legitimate form of creative production. Even if future AI detection systems become substantially more accurate, disputes over authorship may persist because they concern not only provenance but also social values surrounding labor, authenticity, and creativity. These values, however, are not fixed. As artists, commissioners, and communities become more accustomed to generative AI, the social meaning of human authorship and the legitimacy attached to AI-assisted creation may themselves shift over time.

\subsection{Livestreaming as Community-Generated Provenance}
\textit{AI duidu} is one response to this uncertainty. Rather than relying on the artwork itself, the community developed a process-based method for determining the provenance, or origin, of an artwork. While many tools exist today to establish the provenance of digital content, including Meta's AI image labeler \cite{gleeson2024meta}, Hive AI \cite{heikkila2024dod}, and Google SynthID \cite{wiggers2025synthid}, these systems primarily rely on embedded watermarks or metadata generated by participating platforms. None use a full-length recording of the creation process to assess provenance. Notably, our findings suggest that because other forms of process evidence were perceived as relatively easy to fabricate, members of the community turned to livestreams. By providing a real-time record of creation that could not easily be edited, rehearsed, or prepared in advance, and by withholding the commission brief until the livestream began, livestreams became a means of collecting evidence about creative provenance. 

Past work has examined how live recreations of traditionally static forms of artwork can foster engagement \cite{carrera2022watch}, as well as the emerging practice of livestreaming creative work to online audiences \cite{10.1145/3361121, fraser2019sharing}. Researchers have found that creators often adapt their work to make it more engaging for viewers, increasing the performative aspects of creative practice  \cite{fraser2020livestreaming}. Using livestreaming to adjudicate artistic provenance may therefore pressure artists to perform legitimacy through the public display of artistic labor, regardless of whether the disputed work was originally created with AI assistance. Furthermore, this approach assumes that artists can recreate commissioned artwork with a sustained and continuous period of activity. Such expectations may disadvantage artists whose creative process involves extended reflection, multiple iterations, breaks, or other forms of non-linear work. 

Future research should examine how livestream-based forms of provenance verification affect different artists and creative practices. Is it possible that certain artistic workflows are marginalized by the introduction of \textit{AI duidu}, while workflows that lend themselves to public performance, continuous production, and audience engagement are more readily accommodated? More broadly, future work should investigate both the benefits and limitations of using video capture and livestreaming to establish provenance, and whether such approaches are appropriate for community governance, platform moderation, and other contexts in which creative authenticity is contested. 

\subsection{Community Boundary Work and the Protection of Human Labor}
Our findings suggest that resistance to generative AI within the \textit{huiquan} community was not simply a reaction to a new technology, but also a form of collective self-protection and boundary work through which community members distinguished legitimate insiders from outsiders \cite{seering2019moderator, gieryn1983boundary}. Participants described how the value of commissioned artwork was closely tied to assumptions of human labor, creativity, and skill. Consequently, community norms surrounding AI use functioned not only to regulate artistic practice but also to protect the social and economic value of human-made work. Adherence to community norms, along with recognizable artistic development, consistent work practices, and a willingness to discuss and display one's creative process, helped signal legitimate membership within the community. Conversely, accusations of AI use often challenged an artist's status as a legitimate community member, casting them as violating the values that defined belonging. 

At the same time, the boundaries of what constituted ``AI use'' were often ambiguous. Although participants generally agreed that undisclosed AI-generated artwork was problematic, they did not share a common understanding of which specific practices should be considered AI use. Artists and commissioners drew different lines around activities such as reference generation, image editing, ideation, and the incorporation of AI-assisted tools into creative workflows. As a result, disputes often arose not only from concerns about AI itself, but also from competing interpretations of community norms. Community members frequently treated AI use as a binary category despite holding nuanced, and sometimes conflicting, views about which forms of assistance were acceptable. 

These findings resonate with Hwang et al.'s work on AI disclosure in freelance labor, which found that workers struggled not because they resisted disclosure or compliance, but because AI policies were often ambiguous, binary, and disconnected from concrete use cases \cite{hwang2026disclosure}. Similarly, members of the \textit{huiquan} community broadly agreed on the importance of restricting AI use, yet lacked a shared definition of what practices fell within those boundaries. Rather than determining whether an objective rule had been violated, \textit{AI duidu} functioned as a community mechanism through which the meaning of AI use itself was negotiated, contested, and enforced. Through accusations, investigations, livestreamed demonstrations, and collective judgment, community members continually constructed the boundary between acceptable and unacceptable artistic practice. 

While \textit{AI duidu} helped communities address uncertainty surrounding AI authorship, it also revealed an important limitation of community-based enforcement mechanisms. Considerable effort was devoted to identifying and sanctioning suspected rule violators, yet comparatively few mechanisms existed for restoring the reputations of artists who were falsely accused. As a result, even when accusations proved unfounded, the social and economic consequences for accused artists could persist long after a dispute had been resolved.

\subsection{Livestreaming as a Public Adjudication Mechanism}
\textit{AI duidu} also indirectly sought to assess the integrity of the accused artist and determine whether they should remain a trusted member of the community. As discussed in the related work, online communities employ a variety of mechanisms through which moderators or other community members evaluate whether individuals are upholding community standards and should continue to belong. However, \textit{AI duidu} differs from many existing approaches in that moderation is not performed \textit{of} livestreams, but rather \textit{through} livestreams. In this case, livestreams are not the primary content being evaluated; instead, they serve as the mechanism through which adjudication takes place. 

A distinctive aspect of \textit{AI duidu} is that it combines livestreaming with monetary wagering. Unlike prediction markets or sports betting, where spectators wager on the outcome of an event performed by others, the individuals placing the wager are also the parties participating in the dispute. The accused artist and accuser both place money at stake before the livestream begins, creating financial consequences for the outcome of the adjudication. The wager therefore increases the perceived seriousness of the process, signaling confidence in one's position while also discouraging frivolous accusations. Rather than functioning as entertainment or speculation, the wager serves as a mechanism for commitment and accountability within the dispute-resolution process.

Livestreams are a distinctive form of adjudication for several reasons. First, a livestream does not merely provide a record of the creative process, but invites the broader community to observe that process simultaneously. Similar to highly visible and communicative moderation practices described among Twitch moderators \cite{10.1145/3452918.3458796}, and to the public forms of punishment described by Foucault that served to communicate community norms \cite{foucault2012discipline}, the combination of livestreaming and wagering transforms adjudication into a public spectacle with tangible consequences. As the event unfolds, community members are able to communicate with one another, discuss the recreation process, and collectively interpret the evidence in real time. \textit{AI duidu} is therefore not only a means of determining artistic provenance or deciding the outcome of a dispute; it is also a mechanism for communicating, reinforcing, and publicly affirming community norms regarding AI use. 

Further work should explore how livestream-based adjudication can be augmented with tools that support collective sensemaking and consensus building. Prior HCI research has demonstrated how technology can assist online communities in making sense of complex events and coordinating collective judgments  \cite{10.1145/2998181.2998235, 10.1145/3432940}. Additional features that support voting, flagging moments in a livestream for further review, comparing recreated artwork with the original disputed image, or analyzing patterns in audience discussion may help communities reach more informed decisions. Such tools could strengthen the transparency and legitimacy of public adjudication while reducing the effort required to evaluate complex creative disputes.

\section{Conclusion}
Generative AI has introduced new challenges for online art commission communities by undermining members' confidence in the human authorship of digital artwork. Through interviews with participants involved in a Chinese online art commission community (绘圈, huiquan), we examined a novel, community-developed dispute resolution mechanism, the \textit{AI duidu}, that combines livestreaming, wagering, and collective community judgment to adjudicate accusations of generative AI use by artists. Our findings showed that \textit{AI Duidu} shifted authorship verification from the artwork itself to the artist, treating creative labor performed in real time as evidence of provenance. In doing so, it functions not only as a means of resolving disputes but also as a mechanism for negotiating community norms, protecting the perceived value of human creative labor, and defining the boundaries of legitimate artistic participation. 

More broadly, our work suggests that disputes over AI-generated content are not solely problems that can be resolved through improved AI detection systems. Rather, they are also social problems shaped by trust, labor, authenticity, and community governance. As AI-generated content becomes increasingly difficult to distinguish from human-created work, communities may continue to develop new forms of collective verification and adjudication. By documenting the emergence of \textit{AI Duidu}, this case study contributes to a growing understanding of how online communities respond to uncertainty surrounding authorship and authenticity in the age of generative AI.

\section*{Acknowledgments}
The authors used ChatGPT to assist with the language translation and editing, and Claude to assist in making the figures that appear in the paper. All content was reviewed and verified by the authors.

\bibliographystyle{ACM-Reference-Format}
\bibliography{references}

\appendix

\section{Recruitment Message}
\label{appendix:reach-out}
We sent the following message in Chinese to potential participants through Xiaohongshu's direct messaging feature. Its English translation is below.

\begin{quote}
Hello! I am a graduate student in Human-Computer Interaction (HCI) at \anon{the University of Toronto}, and I am currently conducting academic research on the impact of AI technology on digital art communities. Our research focuses on \textit{AI Duidu}, a practice that emerged within the Chinese digital art community in response to suspected AI use in commissioned artwork. We aim to analyze how this practice may affect artists and commissioners.

We would like to invite you to participate in a 60-minute interview to learn about your understanding of \textit{AI Duidu}, your experience taking part in it, and your views on it. Your insights are crucial to this research.

\textbf{Research commitments:}
\begin{itemize}
  \item All interview information will be strictly anonymized and used solely for academic research purposes. This includes both your real identity and your online identities, such as your IDs on social media and commission platforms.
  \item The interview will last approximately 60 minutes.
  \item As a token of our appreciation for your time and contribution, you will receive a stipend of 80 RMB per hour (approximately 16 CAD per hour) upon completion of the interview.
\end{itemize}

If you are interested in this project and willing to share your valuable experiences, we look forward to hearing from you!
\end{quote}

\end{document}